\documentclass[11pt,onecolumn,amssymb,nofootinbib]{revtex4}
\usepackage{amsmath, amsthm, amscd, amssymb}
\usepackage{bm}
\usepackage{bbm}

\begin{document}

\title{\bf Does the Peres experiment using photons test for hyper-complex (quaternionic) quantum theories?}

\author{Stephen L. Adler}
\email{adler@ias.edu} \affiliation{Institute for Advanced Study,
Einstein Drive, Princeton, NJ 08540, USA.}

\begin{abstract}
Assuming the standard axioms for quaternionic quantum theory and
a spatially localized scattering interaction,  the $S$-matrix in quaternionic quantum theory is complex valued, not quaternionic.  Using the standard connections between the $S$-matrix, the forward scattering amplitude for electromagnetic wave scattering, and the index of refraction, we show that the index of refraction is necessarily complex, not quaternionic.  This implies that the recent optical experiment of Procopio et al. based on the Peres
proposal does not test for hyper-complex or quaternionic quantum effects arising within the standard Hilbert
space framework.  Such a test requires looking at near zone fields, not radiation
zone fields.
\end{abstract}

\maketitle

Textbook quantum theory is based on complex numbers of the form $a_0+a_1i$, with $i$ the imaginary unit $i^2=-1$. It has long
been known (see \cite{adlerbook} for a comprehensive review) that an alternative quantum mechanics can be based on the quaternion
or hyper-complex numbers of the form $a_0+a_1i+a_2j+a_3k$, with $i,j,k$ three non-commuting imaginary units obeying the algebra
\begin{align}\label{quat}
&i^2=j^2=k^2=-1~~~,\cr
&ij=-ji=k~~~,\cr
&jk=-kj=i~~~,\cr
&ki=-ik=j~~~.\cr
\end{align}
The standard axioms for quaternionic quantum theory posit [1] a Hilbert space structure with quaternion scalars, a positive
semidefinite inner product, and an inner product-preserving time development. In complex quantum theory time development is generated by a complex unitary  evolution operator $U=e^{iHt}$  with $H$ self-adjoint, and thus $iH$ anti-self-adjoint.  In the quaternionic generalization of quantum theory, this  is replaced by a quaternion unitary  evolution operator $U=e^{\tilde H}$, with the ``quaternionic Hamiltonian''
$\tilde H$ a general quaternion anti-self-adjoint operator.  Because the quaternion algebra is non-commutative, the operator $\tilde H$ cannot be reduced to a self-adjoint operator by multiplication by a fixed quaternion unit.  Quaternionic quantum mechanics is a distinctly
different physical theory from complex quantum mechanics, unlike what is sometimes termed ``quaternion electrodynamics'', which is a rewriting of the Maxwell equations using quaternions in place of the vector calculus, but with the same physical content.

Whether Nature chooses complex or quaternionic quantum theory is ultimately an experimental issue, and Peres \cite{peres} many
years ago proposed a test for quaternionic effects.  His idea was to set up an interferometer with two branches, one
containing materials $A$ and $B$ through which the beam passes, and the other containing $B$ and $A$ in the opposite order.
This would, in principle, test for non-commutativity of the phases $\alpha$ and $\beta$ induced by passage of the beam through
the respective materials $A$ and $B$. Such a non-commutativity  could be present in quaternionic quantum theory, but is zero in complex quantum theory.  Experiments
of this type were carried out with neutrons \cite{neutronref}, with a null result for the phase non-commutativity.  However, in the book \cite{adlerbook}, Adler showed\footnote{ See
Secs. 6.1, 6.3, 7.2, and 8.3 of \cite{adlerbook} for the arguments that the $S$-matrix in quaternionic quantum theory is complex.   For related results for non-Abelian scattering theory, see Soffer \cite{soffer}.} by detailed calculations using the non-relativistic Schr\"odinger equation, and more general arguments using the M\"oller wave operator formalism, that quaternionic
effects in scattering decay asymptotically, with the far-zone scattering amplitude containing only commmuting phases lying in a complex
subalgebra of the quaternions.  So the null results of the neutron scattering experiments do not in fact place useful bounds
on possible quaternionic effects.

Recently, Procopio et al. \cite{pro} reported a version of the Peres experiment using photons, for which the non-relativistic
calculations of \cite{adlerbook} do not apply.  They used photons of wavelength 790 nm, three to four orders of magnitude larger than
 a typical atomic dimension, and so the optical properties of
materials are adequately described by a frequency $\omega$ dependent refractive index $n(\omega)$, which is the zero wave number
limit of the more general frequency and wave number dependent refractive index or dielectric constant \cite{ec}.   In the two
branches of their interferometer, they used optical materials with different optical responses; for $A$ they used a material with
positive refractive index, and for $B$ they used a material with negative refractive index.  For non-dissipative quaternionic
scattering, the phases induced by $A$ and $B$ respectively are assumed to be
\begin{align}\label{phasechange}
\alpha=&e^{i\phi_A^1+j\phi_A^2+k\phi_A^3}~,~~~\cr
\beta =&e^{i\phi_B^1+j\phi_B^2+k\phi_B^3}~~~,\cr
\end{align}
and quaternionic effects (nonzero $\phi_A^{2,3}$ and/or nonzero $\phi_B^{2,3}$) would be signaled by non-commutativity of
the phases $\alpha$ and $\beta$ induced in opposite orders in the two branches of the interferometer.  The experiment gave
a null result, showing identity of the overall phases $\alpha \beta$ and $\beta \alpha$ in the two branches of the interferometer to less than 0.03 degrees of arc. Does this result for photons give a test of quaternionic quantum theory?

To answer this question, we first reexpress the phase $\alpha$ in terms of the refractive index $n$.  For light of wave number $q=\omega/c$ passing through a thickness $w$ of optical material with refractive index $n$, the induced phase is $\alpha=e^{inqw}$.  For a quaternionic
index of refraction $n=n_0+n_1i+n_2j+n_3k$, we thus have $\alpha=e^{(n_0i-n_1+n_2k-n_3j)qw}$.  When the material is non-dissipative
$n_1=0$ and this phase then has the form of Eq. \eqref{phasechange}, with the presence of quaternionic effects signaled by
nonzero values for $n_2$ and $n_3$.  So the question we are asking is whether the index of refraction in quaternionic quantum
theory is complex-valued (only $n_0$ and $n_1$ nonzero) or quaternionic (components $n_2$ and/or $n_3$ present).

The next step is to invoke the Rayleigh relation \cite{gw} between the index of refraction $n(\omega)$ and the coherent
forward scattering amplitude $f(\omega)$,
\begin{equation}\label{lorentz}
n(\omega)=1+ \frac{2 \pi c^2}{\omega^2}N  f(\omega)~~~,
\end{equation}
with  $N$ the number of scattering centers per unit volume.  A concise derivation of this relation
is given in the Fermi lecture notes \cite{fermi}.\footnote{Eq. \eqref{lorentz} holds for $n-1$ small.  When $n$ is of order unity, and the
reduction in the transmitted wave amplitude in the medium by a factor $2/(n+1)$ is taken  into account in a self-consistent way,
the formula becomes \cite{serber}
 \begin{equation}\nonumber
n(\omega)^2=1+ \frac{4 \pi c^2}{\omega^2}N  f(\omega)~~~.
\end{equation}
This change has no effect on our argument, since it still implies that if $f(\omega)$ is complex, then so is $n$.}
Thus the question of whether $n$ can be quaternionic reduces to that of whether
the photon scattering amplitude $f(\omega)$ can be quaternionic.  A detailed analysis of photon scattering is given in the
electrodynamics text of Akhiezer and Berestetskii  \cite{ab}, which gives in Eq. (5.6) the formula\footnote{We make the minor changes
 in notation from \cite{ab} of using a caret to denote unit vectors, and denoting the wave number by $q$ instead of $k$.} for the asymptotic photon
$\vec E $ field for an incident wave propagating along the $z$ direction towards a scatterer at the coordinate origin,
\begin{equation}\label{abformula1}
\surd{2} \vec E= \hat e e^{iqz}+\vec F(\hat n) \frac{e^{iqr}}{r}~~~,
\end{equation}
with $\hat e$ the polarization unit vector and $\hat n$ a unit vector defined by $\vec r=\hat n r$. Denoting by $\hat n_0$
the unit vector along the $z$ axis, Ref. \cite{ab} gives in Eq. (5.10) a formula for $\vec F(\hat n)$, the amplitude for elastic (i.e., non-dissipative)
scattering by a spherically symmetric scatterer,
\begin{equation}\label{abformula2}
\vec F(\hat n)=\frac{2\pi}{iqr} \sum_{jM\lambda} \hat e \cdot \vec Y_{jM}^{\lambda}(\hat n_0)(S_{j\lambda}-1) \vec Y_{jM}^{\lambda}(\hat n) ~~~,
\end{equation}
with $\vec Y_{jM}^{\lambda}$ vector spherical harmonics describing the photon fields, and with $S_{j \lambda}$ the $S$-matrix element in the scattering  channel labeled by $j\lambda$ \big(which is related to the phase shift $\delta_{j \lambda}(\omega)$  by $S_{j \lambda}(\omega)=e^{2i \delta_{j \lambda}(\omega)}$\big).   The question of whether the scattering amplitude $f(\omega)$, which is a component of
$\vec F(\hat n_0)$,
is complex or quaternionic therefore  reduces to that of whether the
$S$-matrix in quaternionic quantum theory is complex or quaternionic.

This question was settled in Sec. 8.3 of \cite{adlerbook} by a very general argument within the standard quaternionic Hilbert
space framework, using the M\"oller wave operator formalism,
which applies for scattering by massless as well as massive particles.  The derivation assumes that the total anti-self-adjoint quaternionic
Hamiltonian $\tilde H$ is the sum of an asymptotic part $\tilde H_0$ and a part $\tilde V$ that is spatially localized, $\tilde H =\tilde H_0 + \tilde V$.
No other  specific assumptions about the structure of $\tilde V$ are made.  The general M\"oller formalism implies that the $S$-matrix commutes
with the asymptotic Hamiltonian,
\begin{equation}\label{comm1}
[\tilde H_0, S]=0.
\end{equation}
We can use the ray representative freedom of quaternionic quantum theory states to choose an energy eigenstate basis for $\tilde H_0$ in which the energy eigenvalues are all of the form $iE$, that
is, the quaternionic phase lies in the complex subspace of the quaternion algebra spanned by 1 and $i$.
Sandwiching Eq. \eqref{comm1} between $\tilde H_0$  eigenstates with eigenvalues $iE$ and $iE^{\prime}$ then implies that the $S$-matrix  element $S(E,E^{\prime})$ is
nonvanishing only when $E=E^{\prime}$, and when $E\neq 0$ the on-energy-shell $S$  matrix commutes with $i$,
\begin{equation}\label{comm2}
[i,S]=0~~~,
\end{equation}
and so is complex.  By the chain of equalities established in the preceding paragraphs, this shows that the index of refraction
in quaternionic  quantum theory is complex, and not quaternionic.

Thus, our conclusion is that the experiment of \cite{pro} does not test for quaternionic effects, unless these are assumed to arise from a
hypothetical theory that does not obey the standard axioms of quaternionic quantum mechanics, for example by having an indefinite inner product.
To test for quaternionic effects arising within the standard Hilbert space framework, one would need an experiment accessing near-zone scattering,
not just the asymptotic or radiation zone scattering amplitude.  The derivation we have given does not show how rapidly near-zone quaternionic effects decay with distance $r$ from
the scattering center.  We conjecture, based on our computations in the non-relativistic case, that even for  massless photons  quaternionic scattering effects
will decay exponentially as $e^{-kr}$, rather than as a power of $r$,  but a detailed calculation using a model for the interaction term $\tilde V$  would be needed to check this.

I wish to thank Angelo Bassi for bringing the paper \cite{pro} to my attention, and to thank  Borivoje Daki\'c, Lorenzo Procopio, Lee
Rozema, and Philip Walther, four  of the authors of \cite{pro}, for a stimulating
email correspondence addressing the issues discussed here.


\begin{thebibliography}{99}

\bibitem{adlerbook}S. L. Adler, {\it Quaternionic Quantum Mechanics and Quantum Fields}, Oxford University Press (1995).
\bibitem{peres}A. Peres, Phys. Rev. Lett.  {\bf 42}, 683 (1979).
\bibitem{neutronref} H. Kaiser, E. George, and S. Werner, Phys. Rev. A {\bf 29}, 2276 (1984); A. G.  Klein, Physica B+C  {\bf 151}, 44 (1988).
\bibitem{soffer} A. Soffer, Phys. Rev. D 29, 1866 (1984).
\bibitem{pro} L. M. Procopio et al., Nature Comm. {\bf 8}, 15044 (2017).
\bibitem{ec} J. Lindhard, Kgl. Danske Videnskab. Selskab, Mat.-fys. Medd. {\bf28}, 8 (1954); for a simple derivation and further references
see H. Ehrenreich and M. H. Cohen, Phys. Rev. {\bf 115}, 786 (1959).
\bibitem{gw} R. G. Newton, Am. J. Phys. {\bf 44}, 639 (1976); Wikipedia article on ``Optical Theorem''.  Both attribute this formula to
Lord Rayleigh.
\bibitem{fermi} E. Fermi,   {\it Nuclear Physics}, University of Chiciago Press, 1950 Revised Edition, pp. 201-202.  A derivation is also given in
the Wikipedia article on ``Optical Theorem''.
\bibitem{serber}  R. H. Serber, {\it Serber Says: About Nuclear Physics},  World Scientific Lecture Notes in Physics,
Vol. 10 (1987), pp. 50-51, derivation attributed to G. C. Wick; S. L. Adler, unpublished notes (late 1960s).
\bibitem{ab} A. I. Akhiezer and V. B. Berestetskii, {\it Quantum Electrodynamics}, Interscience Publishers (1965), pp. 39-40.
\end{thebibliography}
\end{document}